\documentclass[12pt,preprint]{aastex} 

\def\gtorder{\mathrel{\raise.3ex\hbox{$>$}\mkern-14mu
             \lower0.6ex\hbox{$\sim$}}} 
\def\ltsima{$\; \buildrel < \over \sim \;$}
\def\simlt{\lower.5ex\hbox{\ltsima}}
\def\gtsima{$\; \buildrel > \over \sim \;$}
\def\simgt{\lower.5ex\hbox{\gtsima}} 

\begin{document} 


\title{Real-Time Detection and Rapid Multiwavelength Follow-up Observations 
of a Highly Subluminous Type II-P Supernova from the Palomar Transient 
Factory Survey}


\author{Avishay Gal-Yam}
\affil{Department of Particle Physics and Astrophysics, Faculty of Physics, The Weizmann
Institute of Science, Rehovot 76100, Israel}
\email{avishay.gal-yam@weizmann.ac.il}

\author{Mansi M. Kasliwal}
\affil{Cahill Center for Astrophysics, California Institute of Technology, Pasadena, CA 91125, USA}


\author{Iair Arcavi, Yoav Green, Ofer Yaron, Sagi Ben-Ami, Dong Xu, Assaf Sternberg}
\affil{Department of Particle Physics and Astrophysics, Faculty of Physics, The Weizmann
Institute of Science, Rehovot 76100, Israel}

\author{Robert M. Quimby, Shrinivas R. Kulkarni, Eran O. Ofek\footnotemark[1], Richard Walters}
\affil{Cahill Center for Astrophysics, California Institute of Technology, Pasadena, CA 91125, USA}

\footnotetext[1]{Einstein Fellow}

\author{Peter E. Nugent, Dovi Poznanski\footnotemark[1]}
\affil{Computational Cosmology Center, Lawrence Berkeley National Laboratory, 1 Cyclotron Road, Berkeley,
CA 94720, USA}

\author{Joshua S. Bloom, S. Bradley Cenko, Alexei V. Filippenko, Weidong Li, J. Silverman}
\affil{Department of Astronomy, University of California, Berkeley, CA, 94720-3411, USA}

\author{Emma S. Walker}
\affil{Scuola Normale Superiore, Piazza dei Cavalieri, 7, 56126 Pisa, Italy}

\author{Mark Sullivan, K. Maguire}
\affil{Department of Physics (Astrophysics), University of Oxford, Keble Road, Oxford, OX1 3RH, UK}

\author{D. Andrew Howell}
\affil{Las Cumbres Observatory Global Telescope Network, Goleta,
CA 93117, USA, and Department of Physics, University of California, Santa Barbara,
CA 93106-9530, USA}

\author{Paolo A. Mazzali}
\affil{Scuola Normale Superiore, Piazza dei Cavalieri, 7, 56126 Pisa, Italy; INAF – Osservatorio Astronomico, vicolo dell’Osservatorio, 5, I-35122 Padova, Italy; Max-Planck Institut f¨ur Astrophysik, Karl-Schwarzschild-Str. 1, D-85748 Garching, Germany}

\author{Dale A. Frail}
\affil{National Radio Astronomy Observatory, P.O. Box O, Socorro, NM 87801, USA}

\author{David Bersier, Phil A. James}
\affil{Astrophysics Research Institute, Liverpool John Moores University, Twelve Quays House, Egerton Wharf, Birkenhead CH41 1LD, UK}

\author{C. W. Akerlof}
\affil{Randall Laboratory of Physics, University of Michigan, 450 Church Street, Ann Arbor, MI 48109-1040, USA}

\author{Fang Yuan}
\affil{Research School of Astronomy and Astrophysics, The Australian  
National University, Cotter Road, Weston Creek, ACT 2611,Australia}

\author{Derek B. Fox}
\affil{Department of Astronomy and Astrophysics, Pennsylvania State University, 525 Davey Laboratory, University Park, PA 16802, USA}

\author{Nicholas Law}
\affil{Dunlap Institute for Astronomy and Astrophysics, University of Toronto, 50 St. George Street, Toronto
M5S 3H4, Ontario, Canada}

\and

\author{Neil Gehrels}
\affil{NASA Goddard Space Flight Center, Greenbelt, MD 20771, USA}



\begin{abstract} 
The Palomar Transient Factory (PTF) is an optical wide-field
variability survey carried out using a camera with a 7.8 square degree
field of view mounted on the 48-in Oschin Schmidt telescope at Palomar
Observatory. One of the key goals of this survey is to conduct
high-cadence monitoring of the sky in order to detect optical
transient sources shortly after they occur. Here, we describe the
real-time capabilities of the PTF and our related rapid
multiwavelength follow-up programs, extending from the radio to the
$\gamma$-ray bands.  We present as a case study observations of the
optical transient PTF10vdl (SN 2010id), revealed to be a very young
core-collapse (Type II-P) supernova having a remarkably low
luminosity.  Our results demonstrate that the PTF now provides for
optical transients the real-time discovery and rapid-response
follow-up capabilities previously reserved only for high-energy
transients like gamma-ray bursts.
\end{abstract} 




\section{Introduction} 

The study of cosmic explosions, the energetic events marking the
deaths of stars, is a unique subfield of astrophysics in the sense
that its research subjects are variable sources on timescales much
shorter than human lifetimes. Furthermore, many physical aspects of
these events can only be revealed by observations conducted shortly
after explosion.  An extreme case are gamma-ray bursts (GRBs; e.g.,
Piran 2004; Woosley \& Bloom 2006; Nakar 2007), where the most
energetic radiation is emitted within a few seconds after the
explosion.  The study of GRBs has seen much progress following the
launch of the {\it Swift} satellite, enabling rapid multiwavelength
follow-up observations of GRBs.

Investigation of optical transients has thus far lagged behind in this
respect. Rapidly detecting optical transients from among the multitude
of constant (and variable) sources visible in the data stream from
wide-field optical surveys has proved challenging for existing
systems. Initial progress in securing very early-time observations of
optical transients was made with relatively smaller surveys such as
the Lick Observatory Supernova Search using the Katzman
Automatic Imaging Telescope (KAIT; e.g., Ganeshalingam et al. 2010)
and the Texas Supernova Survey using the ROTSE telescopes (e.g.,
Quimby et al. 2007), or due to chance coincidences such as supernovae
(SNe) occurring in fields being monitored due to previous supernova
(SN) explosions (e.g., Stritzinger et al. 2002; Soderberg et
al. 2008).

One of the main science goals of the Palomar Transient Factory
(PTF\footnote{http://www.astro.caltech.edu/ptf/}) wide-field
variability survey (Rau et al. 2009; Law et al. 2009) is to remedy
this situation, and to facilitate real-time discovery and prompt
monitoring of optical transients. Following recent upgrades to the
computing infrastructure and software that now process all PTF data in
real time and disseminate internal alerts about transients, and
finalization of the operational aspects required to make use of these
discoveries, the PTF is now a real-time optical variability
experiment. Target-of-opportunity (ToO) programs are in place to
conduct rapid multiwavelength follow-up observations of PTF
transients.

Here, we describe the PTF operational setup enabling real-time
discovery and rapid ToO observations of optical transients. We
describe in some detail a specific example, the discovery and
multiwavelength follow-up campaign of the optical transient
PTF10vdl. Our observations of this remarkably subluminous Type II-P SN
demonstrate the scientific potential of data collected shortly after
explosion.

\section{PTF Real-Time Detection and Prompt Monitoring}

PTF images (currently $11 \times 8.4$~megapixels per exposure, or
184\,MB per full array readout) are recorded on computers inside the
48-in Schmidt telescope dome at Palomar Observatory, and are then
rapidly transmitted to the National Energy Research Scientific
Computing Center (NERSC) at Lawrence Berkeley National Laboratory
(LBNL).  A series of software components is then run in sequence
(Nugent et al. 2011, in preparation), including image detrending,
astrometry, image subtraction with respect to deep references produced
from coadditions of previous PTF data, and object detection in the
difference images.

All of these data are loaded into a database upon which a series of
quality cuts to reject artifacts is performed, followed by a rough
classification into four groups by the {\tt Oarical} software. 
These groups include candidate SNe,
sources in galactic nuclei, variable point sources, and moving
objects. Within $\sim 40$ min from image acquisition, internal alerts
are circulated, informing interested collaboration members of recent
discoveries and classifications.  This selection process will be
described in more detail in a forthcoming publication (Bloom et
al. 2011, in preparation).

Once the automated discovery pipeline runs to completion, several
additional processes begin. First, follow-up observations for the most
promising candidates (candidate SNe or sources near known nearby
galaxies) are robotically triggered.  Optical imaging is scheduled at
the robotic 60-in (1.5\,m) telescope (P60; Cenko et al. 2006) at
Palomar Observatory, while an infrared follow-up request is sent to
PAIRITEL\footnote{http://www.pairitel.org/}.  Next, manual scanning of
recent discoveries is conducted. Taking advantage of the 10\,hr time
difference between Israel and California, a duty astronomer at the
Weizmann Institute monitors recent discoveries from Palomar during
Israeli daytime, and can manually trigger additional follow-up
observations if desired (see below).  At the end of each night, good
candidates are sent to be screened by the citizen-science ``supernova
zoo'' program (Smith et al. 2011), while custom database queries are
run by collaboration members with specific focused interests (local
Universe transients, e.g., Kasliwal et al. 2011; Type Ia SNe, e.g.,
Cooke et al. 2011; SN shock breakouts, e.g., Ofek et al. 2010).
Objects tagged by the SN zoo volunteers are also manually reviewed by
the Weizmann duty astronomer. Upon discovery of an object deemed of
special interest, follow-up programs are triggered. 

Between January 1 - April 1, 2011, there were 70 nights 
during which the survey was running for at least part of each night.  
During those nights, there were 732 automated alerts for extragactic
transients (so about 10 per night, on average), of which 296 were 
considered likely SNe, 208 likely AGN (mostly from known sources) and
228 were nuclear events of undetermined nature. 
Weizmann Duty astronomers saved 172 events as
high-priority SN candidates for follow-up; of those 97 ($56\%$) were
spectroscopically followed-up, leading to the discovery of 79 SNe 
($46\%$, of which 64 were of type Ia). The events observed spectroscopically
but not determined to be SNe were CVs or other variable stars (9), 
four galaxy spectra (likely resulting from target SNe fading away, errors in
spectroscopic acquisition or spurious detections), four spectra which
were of too low S/N to identify the nature of the target, and one asteroid.     
During spring/summer time the PTF discovery rates are substantially higher,
due to a combination of better weather at Palomar and the survey strategy,
however, the realtime procedures described here were only perfected 
during summer 2010, and we therefore provide statistics based on the
above period.

Rapid follow-up programs designed to monitor PTF discoveries (mostly
in ToO mode) are in place. A partial list of currently active programs
includes access to the Wise Observatory 1-m and 18-in telescopes,
which, with the benefit of the 10\,hr time difference with respect to
Palomar, can provide rapid ``same-night'' confirmation imaging (see
below). Additional ToO photometry can be obtained by the Las Cumbres
Observatory Global Telescope network (LCOGT), while continued optical
monitoring of SNe is conducted using the 0.75\,m Katzman Automatic
Imaging Telescope (KAIT; Filippenko et al. 2001) at Lick Observatory,
and at the Liverpool Telescope (LT) in La Palma. ToO spectroscopic
programs, as well as many classically scheduled nights, provide access
to the 10\,m Keck telescopes, the 5\,m Hale telescope at Palomar
Observatory, the 3\,m Shane telescope at Lick Observatory, the 8.1\,m
Gemini North and South telescopes, the 8.2\,m Very Large Telescope
(VLT) units at the European Southern Observatory (ESO) site on
Paranal, the 9.2\,m Hobby-Eberly Telescope (HET) at McDonald
Observatory, the 3.5\,m Telescopio Nazionale Galileo (TNG; service
queue mode) in La Palma (Spain), and the University of Hawaii 2.2\,m
telescope on Mauna Kea. There are also radio follow-up programs with
the Extended Very Large Array (EVLA\footnote{The Very Large Array is
  operated by the National Radio Astronomy Observatory, a facility of
  the National Science Foundation operated under cooperative agreement
  by Associated Universities, Inc.}), the Allen Telescope Array (ATA),
and the Combined Array for Research in Millimeter-wave Astronomy
(CARMA) facility (e.g., Carpenter 2010; Corsi et al. 2011). 
Ultraviolet (UV), X-ray, and $\gamma$-ray observations
are obtained with the {\it Swift} mission. These resources provide a
high probability that multiwavelength follow-up observations for an
interesting PTF target can be obtained on short timescales (hours). We
exemplify this capability in the next section using observations of
PTF10vdl. Other recent examples include PTF10fqs (Kasliwal et
al. 2011), PTF10hjz (Frail et al. 2011, in preparation), and PTF10iya
(Cenko et al. 2011).
  
\section{Observations of PTF10vdl}

\subsection{Detection Timeline}

\begin{figure}
\includegraphics[width=14cm]{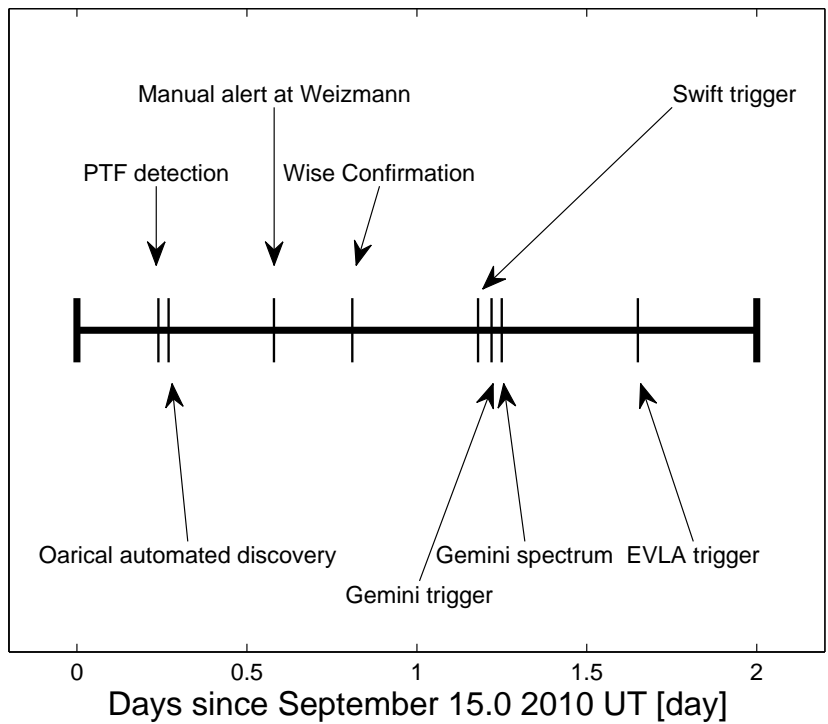}
\includegraphics[width=14cm]{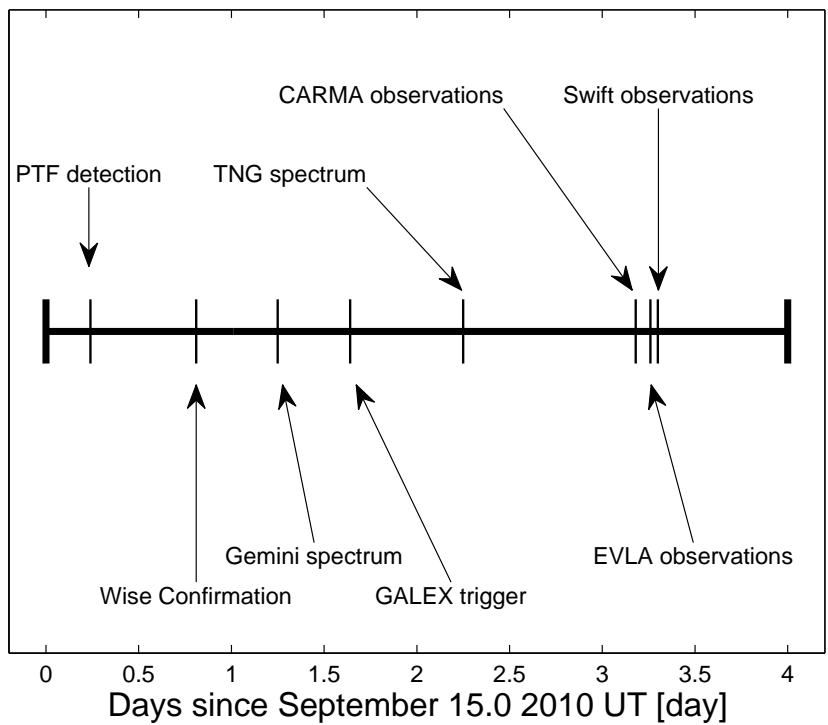}
\caption{Top: Time line of activities following the initial detection of
  PTF10vdl at Palomar. Bottom: Multiwavelength observations of PTF10vdl; 
see text for details.}
\label{timeshortfig}
\end{figure}

PTF10vdl was detected at the 48-in Oschin Schmidt telescope (P48) on
2010 September 15.24 (UT dates are used throughout), and was not
detected in previous images obtained by the survey up to 3.9\,days
earlier. PTF10vdl is located at $\alpha = 23^{\rm h}05^{\rm m}48^{\rm
  s}.88$, $\delta = +03^{\circ}31'25''.5$ (J2000).  The survey data
were downlinked to the NERSC computing center at LBNL and were
processed in real time. The object was discovered by the automated
{\tt Oarical} software shortly (43 min) after detection, and an internal
notification was broadcast. The Weizmann astronomer on duty (Y.G.)
scanned the automated discovery reports as they came in during Israeli
daytime and sent an alert concerning PTF10vdl and other recent
discoveries of interest on September 15.58. The association of this
optical transient with a nearby host galaxy (NGC 7483 at redshift
$z=0.016475$; Springob et al. 2005, via NED) focused our attention on
this event.  The timeline of activities following that alert is
presented in Fig.~\ref{timeshortfig} (top), and included obtaining a
confirmation image of the object using the Wise Observatory 1\,m
telescope, followed by triggering approved PTF ToO programs using the
{\it Swift} satellite, the EVLA radio telescope, and spectroscopy
using the Gemini-North and TNG telescopes.

\subsection{Follow-up Observations}

\begin{figure}
\includegraphics[width=1\textwidth]{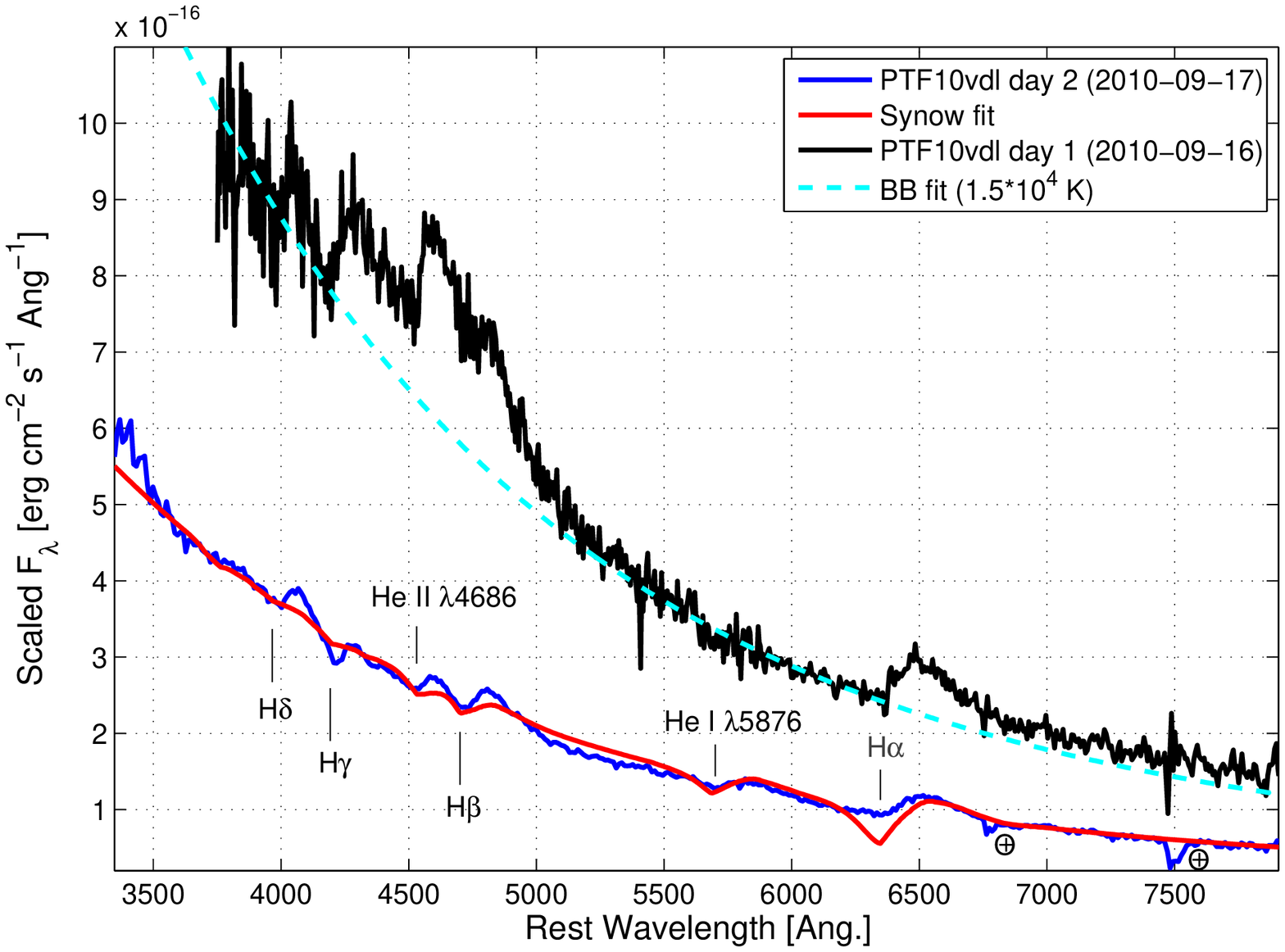}
\caption{Early-time spectra of PTF10vdl. The Gemini spectrum (top)
  exhibits a very blue continuum (a black-body curve with $T_{\rm BB}
  = 1.5 \times 10^4$\,K is shown for reference; cyan dashed curve),
  with superposed line emission dominated by hydrogen and helium.  A
  higher signal-to-noise ratio spectrum of PTF10vdl obtained with the
  3.5\,m TNG about 1 day after the initial Gemini spectrum was fit
  using the SYNOW parametric spectral synthesis code; line
  identifications are marked (using the same $T_{\rm BB}$ as
  above). Note the strong He~II emission, suggesting a very hot
  photosphere. The spectra indicate that this is a very young SN~II,
  with very few similar spectra on record (e.g., for SN 2006bp, Quimby
  et al. 2007).}
\label{earlyspecfig}
\end{figure}

After the rapid spectroscopic confirmation of PTF10vdl
(Fig.~\ref{earlyspecfig}), we obtained multiwavelength observations of this
object. {\it Swift} UV and X-ray observations were rapidly scheduled,
but could not be carried out until Sep. 18 (Kasliwal et al. 2010a) due
to Sun-angle constraints (XRT cooling). No X-rays or $\gamma$-rays
were detected; optical/UV observations were reduced with custom
scripts based on the prescription of Poole et al. (2008) and are shown
in Fig. 5.  {\it GALEX} UV spectroscopy was triggered, but could not
be carried out due to the proximity of PTF10vdl to the UV-bright star
$\beta$~Psc.

CARMA and EVLA radio observations were obtained on Sep. 18 as well,
yielding upper limits (Carpenter et al. 2010; Kasliwal et al. 2010b;
Fig.~\ref{timeshortfig}, bottom). The EVLA observations were made on Sep. 18.26
at a center frequency of 8.46\,GHz; we did not detect radio emission
from the position of PTF10vdl down to a 3$\sigma$ limit of
105\,$\mu$Jy. The mean frequency of the CARMA observations was
97.5\,GHz with a total bandwidth of 8\,GHz.  PTF10vdl was not
detected, with a measured flux density of $0.13 \pm 0.13$\,mJy.

\begin{figure}
\includegraphics[width=1\textwidth]{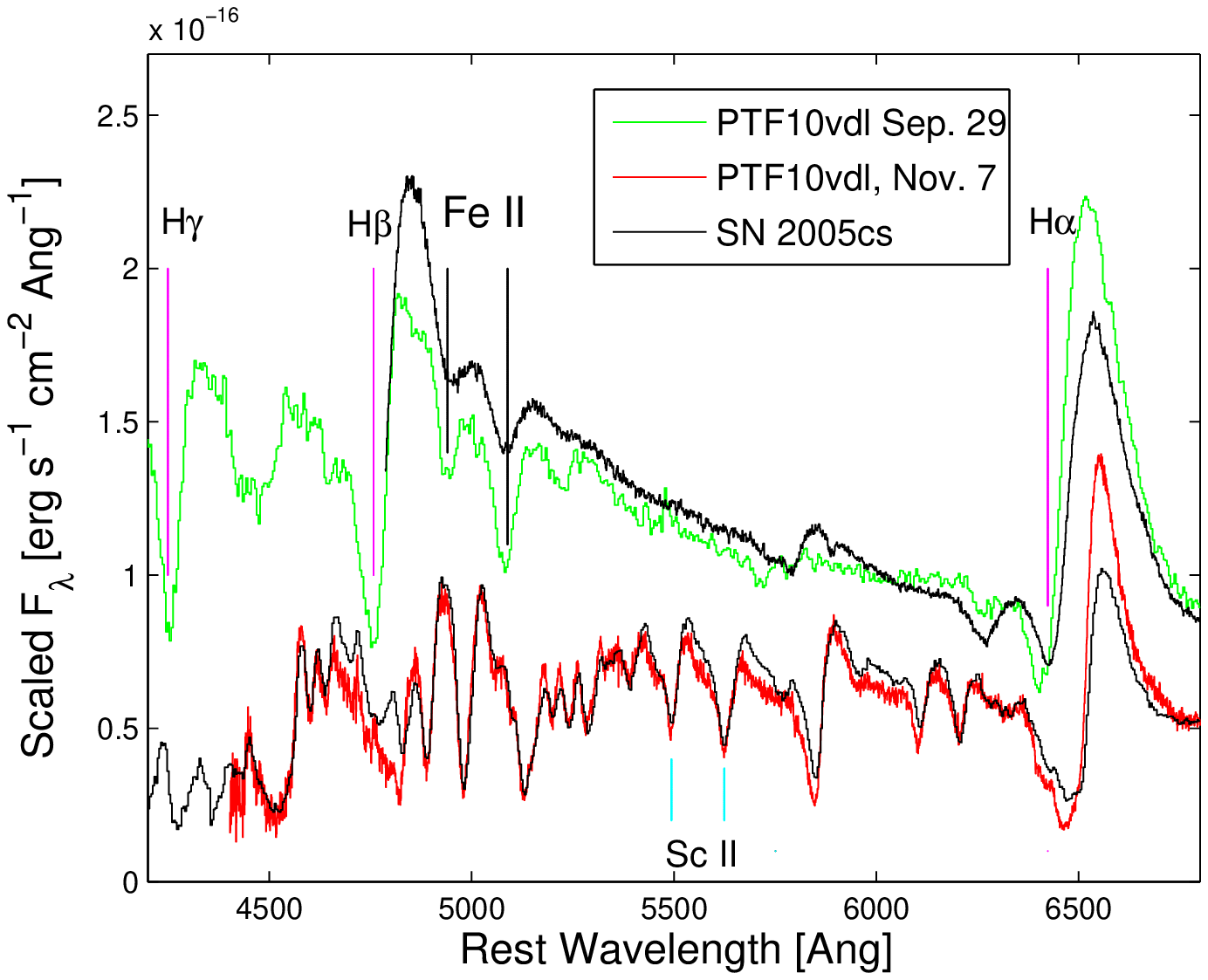}
\caption{Spectral evolution of PTF10vdl. We show two later spectra of
  this event, covering the range 2--8 weeks after explosion; a full
  description will be presented in a future publication. The first
  spectrum, obtained with the Low-Resolution Spectrograph (LRS)
  mounted on the HET, is the first to show a typical spectrum for a
  SN~II during the plateau phase (e.g., Filippenko 1997). Expansion
  velocities derived from minima of developing Fe lines are relatively
  low ($\sim 4700$\,km\,s$^{-1}$).  As the object ages on the plateau,
  as seen in the second spectrum obtained with the DEIMOS spectrograph
  mounted on the Keck-II telescope, weaker lines of other elements
  develop; velocities measured from the Sc~II lines are very low
  ($\sim 1800$\,km\,s$^{-1}$).  The velocity evolution we measure is
  consistent with the trends seen in other low-luminosity SNe~II-P
  (e.g., Pastorello et al. 2004, 2009); direct comparison is shown
  with spectra of SN 2005cs (black lines) obtained 6 days (top) and 33
  days (bottom) after peak (from Pastorello et al. 2009). 
  }
\label{latespecfig}
\end{figure}

\begin{figure}
\includegraphics[width=1\textwidth]{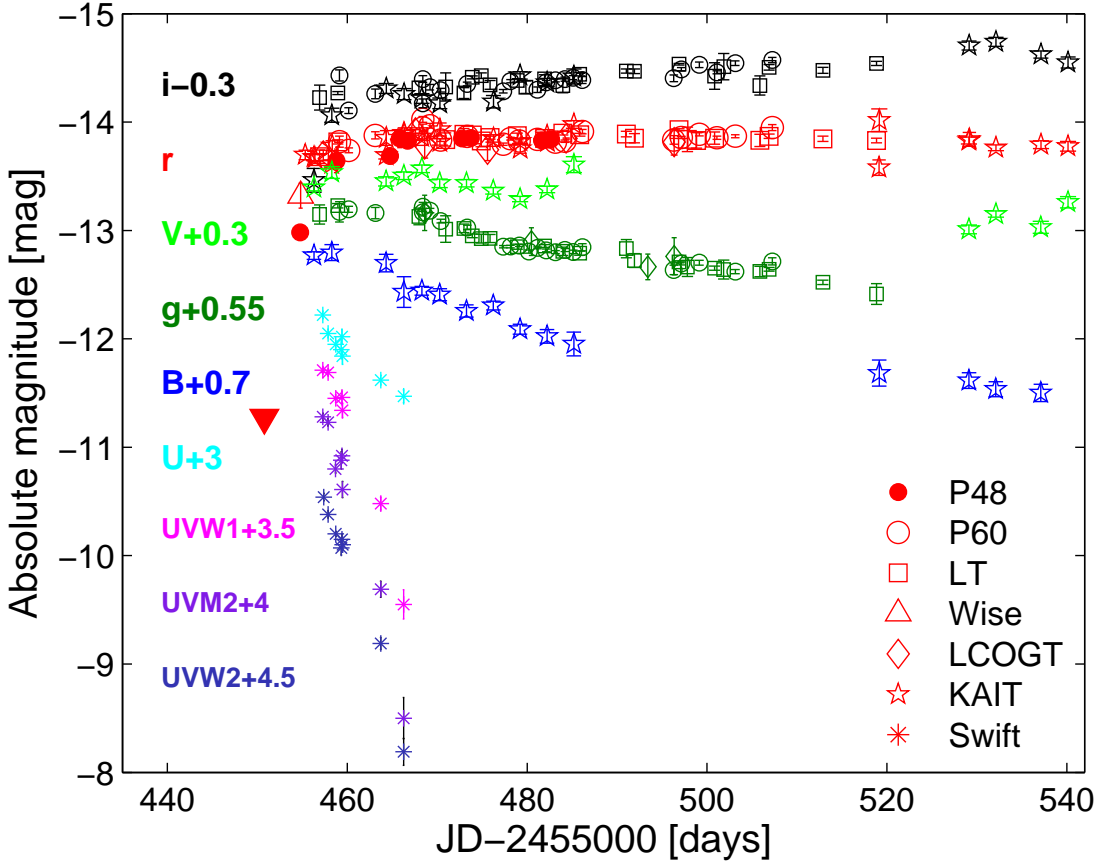}
\caption{The light curve of PTF10vdl. P48 points obtained from the
  survey data using the ``white glove'' photometry pipeline (Poznanski
  et al. 2011, in preparation) daily averages are shown as red filled
  circles, and the last nondetection as a downward-facing
  triangle. Ground-based $BgVri$ and {\it Swift} $U$, $UVW1$, $UVM2$,
  and $UVW2$ measurements are shown in blue, dark green, light green,
  red, black, cyan, magenta, purple, and indigo, respectively; the
  data source is indicated by the marked shape. KAIT $I$ was
  transformed to the SDSS $i$-band grid to match $i$-band data from
  other sources; KAIT unfiltered and all $r/R$-band data were offset
  to align with the P60/LT data calibrated onto the SDSS $r$-band
  zeropoint. All data are reported on the absolute scale assuming a
  distance modulus of $\mu=32.86$\,mag (from NED). For clarity, $i$,
  $V$, $g$, $B$, $U$, $UVW1$, $UVM2$, and $UVW2$ data are offset by
  $-0.3$, 0.3, 0.55, 0.7, 3, 3.5, 4, and 4.5\,mag, respectively. The
  rapid rise is followed by a constant plateau, the hallmark of
  SNe~II-P. The absolute $r$-band plateau luminosity is remarkably
  faint, $M_{R} = -13.85$ mag, comparable only to SN 1999br
  (Pastorello et al. 2004).}
\label{LCfig}
\end{figure}

Rapid ToO spectroscopic observations with Gemini-North were conducted
with the Gemini-N Multi-Object Spectrograph (GMOS; Hook et al. 2004),
configured with a $1''$ slit and the R400 and B600 grisms (720\,s and
520\,s, exposures, respectively), yielding one of the earliest spectra
obtained of any SN (Fig.~\ref{earlyspecfig}).  Shortly thereafter we
obtained service queue-mode spectroscopy with the 3.5\,m TNG using the
Device Optimized for the LOw RESolution (DOLORES) spectrograph,
configured with the LR-B grism and a $1''$ slit. Two exposures of
1200\,s each were taken at the parallactic angle (Filippenko 1982)
under good conditions with $1.4''$ seeing.  Both spectra were reduced
within IRAF in a standard manner. In both spectra, emission lines of H
and He are superposed on a very blue continuum, indicating that this
was a young Type II SN. The remarkably strong, high-excitation He~II
lines are indicative of the initially high temperatures. Additional
spectroscopic monitoring was conducted with the 9.2\,m HET, 3\,m Lick,
5\,m Palomar, and 10\,m Keck telescopes
(Fig.~\ref{latespecfig}). Photometric monitoring was carried out using
the 1\,m Wise, 2\,m LT, P60, KAIT, and 2\,m LCOGT Faulkes-North
telescopes. The resulting light curve (Fig.~\ref{LCfig}) shows that
PTF10vdl is a remarkably underluminous SN~II-P.

Digital copies of our data can be downloaded directly from the 
Weizmann Institute of Science Experimental Astrophysics Spectroscopy System
(WISEASS\footnote{http://www.weizmann.ac.il/astrophysics/wiseass/}, Yaron et al. 2011, 
in preparation).



\section{Results}

The combination of the PTF real-time analysis pipeline with
rapid-response follow-up programs enables us to launch multiwavelength
monitoring campaigns of optical transients, such as PTF10vdl, within
hours of discovery. Since the observations of this particular object
are still ongoing, a full discussion will be presented in future
publications. Here, we report on the aspects that are unique to the
rapid follow-up campaign we conducted.

Our dataset includes a remarkable early-time spectrum (within 2\,days
of discovery; Fig.~\ref{earlyspecfig}), rarely obtained for SNe~II in
the past. It is similar to observations of a handful of other SNe~II
observed at comparably early ages.  In particular, thermally excited
He~II lines are strongly detected, and they disappear within a few
days after the explosion as the expanding photosphere cools\footnote{We note 
that line identification in very early SN II spectra, and in particular,
the relative contributions of He II, N II and O II, is a matter
of some debate (e.g., Quimby et al. 2007; Baron et al. 2007, Dessart et al. 2008);
further analysis of our data in conjunction with detailed models
is encouraged.}. To the
best of our knowledge, these are the earliest spectra ever obtained of
an underluminous SN II-P. For example, spectroscopic coverage of SN
2005cs in M51, the best-observed member of this subclass, probably
began about a day later than our earliest spectrum, and He~II lines 
are not seen (Pastorello et al. 2009).

Our photometric coverage of this event extends from the far UV to the
$i$ band, tracing the rise of the light curve onto the plateau with
excellent temporal sampling. We measure an $r$-band plateau absolute
magnitude of $M_r = -13.85$, which is remarkably faint; only a single
other SN~II-P, SN 1999br (Pastorello et al. 2004), has a comparable
plateau luminosity ($M_V = 13.76$\,mag). Linear decline rates measured
in the $UVW2$, $UVM2$, $UVW1$, $U$, $B$, $g$, $V$, $r$, and $i$ bands
are 0.2509, 0.2995, 0.2375, 0.0743, 0.0191, 0.0149, 0.0120, $-0.0008$, and
$-0.0043$\,mag\,d$^{-1}$, respectively. Bands bluer than $g$ show
a rapid decline starting with our earliest data points, while the
$i$ band exhibits a slow rise during the initial phase of the plateau.
 
\begin{figure}
\includegraphics[width=1\textwidth]{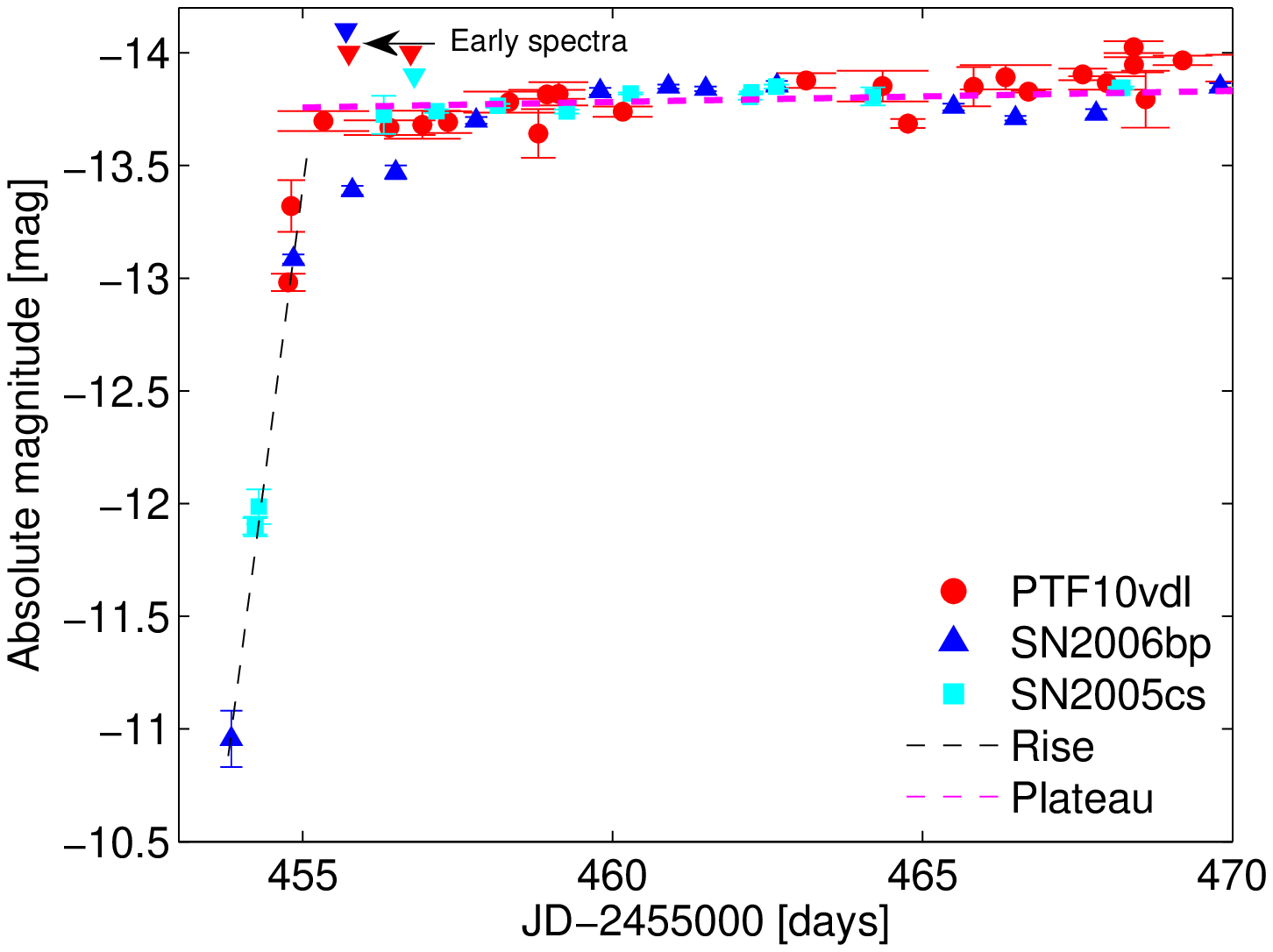}
\caption{Early-time light curves of SNe~II-P. We compare our $R$-band
  light curve of PTF10vdl (red) with those of the normal SN II-P
  2006bp (daily averages from Quimby et al. 2007; blue) and the
  underluminous SN 2005cs (Pastorello et al. 2009; cyan). We have
  scaled the flux to align the plateau luminosities (magenta dashed
  line). As shown by Pastorello et al. (2009), the rapid rise phase of
  SNe 2005cs and 2006bp is consistent with a single linear slope (in
  magnitude vs. time; black dashed line), and we have adjusted the
  estimated explosion date of PTF10vdl (which is not as well
  constrained as those of the other events) so that the earliest
  observations tracing the rise fall on the same slope. It appears as
  if the underluminous events, PTF10vdl and SN 2005cs, transition
  sharply from a rise to a plateau, compared to the more luminous SN
  2006bp. Additional events (and even better data) are required to
  determine if this is a general trend.  The dates of the earliest
  spectra for these three events (from Pastorello et al. 2006, Quimby
  et al. 2007, and this work) are marked with triangles having colors
  matching the figure legend.}
\label{LCzoomfig}
\end{figure}

The only two other SNe~II-P observed this early with well-defined
plateau light curves (c.f. the peculiar SN 1987A)
were SN 2005cs (underluminous; Pastorello et al. 2009) and SN 2006bp
(normal; Quimby et al. 2007). Comparison with PTF10vdl shows a possible
trend: it seems that both underluminous events (SN 2005cs and
PTF10vdl) have a fast rise with a sharp onset of the flat plateau,
while the more luminous SN 2006bp has a more gradual transition, with
the fast rise leveling onto the plateau over a http://arXiv.org/help/faq/texlivespan of a few days
(Fig.~\ref{LCzoomfig}; see also Fig. 3 of Pastorello et
al. 2009). Future early (and dense) monitoring of additional SNe~II-P
should clarify whether there is a correlation between the plateau
luminosity and the morphology of the early-time light curve;
presumably such differences, if they exist, reflect the physical
structure of the outermost layers of the exploding star. At these
early times, line emission is weak and the $r$-band flux is dominated
by continuum emission on the Rayleigh-Jeans tail of the Planck curve,
so its evolution should mirror that of the bolometric luminosity.
 
Type II-P SNe have been shown to be promising ``standard candles''
using an empirical relation between the photospheric velocity
(measured from weak absorption lines, e.g., of Fe~II), a color term
(that traces dust extinction and intrinsic variance), and the
luminosity (Hamuy \& Pinto 2002; Nugent et al. 2006; Poznanski et
al. 2009, 2010).  While low-luminosity events such as PTF10vdl are
expected to be rare in flux-limited cosmological SN samples, it is
instructive to examine whether this event follows the general
relation. Using the parameters from Poznanski et al. (2009), we find
that PTF10vdl is a significant outlier, $\sim 1$\,mag away from the
benchmark set by higher luminosity SNe~II-P. Its velocity on the
plateau (1800\,km\,s$^{-1}$; Fig. 5) translates into a positive
luminosity correction of $\sim 2$\,mag, with an additional 0.25\,mag
from its color correction. However, its plateau luminosity is more
than 3\,mag fainter than that of the average SN~II-P. We note that SN
2005cs (which is somewhat brighter than PTF10vdl, but still considered
among the low-luminosity SN~II-P group; Pastorello et al. 2009) does
follow the cosmological relation. Even if some fainter SNe~II-P like
PTF10vdl violate the correlations established at higher luminosities,
this is unlikely to impede future efforts to use SNe~II-P as
cosmological beacons, as such faint objects are rare in flux-limited
samples. On the other hand, this apparent anomaly motivates further
studies of low-luminosity SNe~II-P and continued investigation of the
physics driving observed correlations (e.g., Kasen \& Woosley 2009).
The combination of early spectroscopic observations and UV/optical
light curves makes PTF10vdl an attractive target for such studies, as
well as for comparison with detailed explosion models (e.g., Baron et
al. 2004; Dessart et al. 2008).

Our {\it Swift} observations resulted in solid detections of PTF10vdl
in all UV filters (Kasliwal et al. 2010a), and a nondetection in
X-rays. CARMA and EVLA observations also resulted in upper limits
(Carpenter et al. 2010; Kasliwal et al. 2010a). It is interesting to
note that both {\it Swift} and EVLA observations were delayed by $\sim
48$\,hr due to technical limitations that should generally not apply;
thus, future PTF events should be observed even sooner, increasing the
exciting prospect of X-ray, UV, and radio detection of prompt,
nonthermal radiation from cosmic explosions detected by PTF.

Note that the Lick Observatory Supernova Search (LOSS; Filippenko et
al. 2001) discovered the same object $\sim 1$\,day after our PTF
discovery (Lin et al. 2010), but CBET procedures prevented the
dissemination of this information for several days, until it was
announced as SN 2010id. It thus seems that high-cadence observations
of nearby galaxies by LOSS should also be able to contribute to future
discoveries at very early times.

\section{Discussion and Conclusions}

Early detection and rapid follow-up observations, as demonstrated with
PTF discoveries, are interesting for a number of reasons.  The first
electromagnetic manifestation of the explosions of massive stars as
core-collapse SNe would be the process of the outgoing shock emerging
from the stellar envelope (so-called ``shock-breakout''), followed by
a period of rapid radiative cooling. PTF optical data alone can now be
used to detect such explosions during the first hours after they
occur, and to trigger space-based follow-up observations in the UV and
X-rays. At least for exploding red supergiants, the few-hour timescale
should allow observations of the tail (at least) of the shock-breakout
process as well as the subsequent cooling, enabling measurement of the
radius of the exploding star and determination of its atmospheric
composition (e.g., Rabinak \& Waxman 2011), and perhaps even
constraining shock-breakout physics (e.g., Katz, Budnik, \& Waxman
2010; Nakar \& Sari 2011). Such data would be extremely valuable to
guide and constrain massive-star evolution models. It is not yet clear
whether the observations of PTF10vdl presented here would prove useful
in this respect. Very early
observations of thermonuclear SNe~Ia (for recent PTF examples, see
Cooke et al. 2011; Howell et al. 2010; Kasliwal et al. 2010c) would
provide constraints on the progenitors of these important cosmic
distance estimators (e.g., Kasen 2010).

Several more exotic scenarios predict fast transients on $\sim 1$\,day
timescales (e.g., compact-object mergers, Metzger et al. 2010;
accretion-induced collapse of a white dwarf to a neutron star, Metzger
et al. 2009). If these transients emit enough red light to be detected
by the PTF $r$-band survey, rapid follow-up observations would be
critical for studying their properties during their fleetingly short
luminous phases.

An important aspect common to the processes mentioned above (SNe,
compact-object mergers) is that they are expected to be leading
sources of gravitational waves in the advanced LIGO/VIRGO bands, as
well as potential sources of cosmic neutrinos.  A narrow timing window
based on coincidence with PTF optical detections would allow searching
for signal over a significantly shorter stream of background noise,
thereby reducing the minimal detection threshold and increasing the
effective sensitivity of existing detectors (e.g., Kochanek \& Piran
1993; Cowen, Franckowiak, \& Kowalski 2010).  While PTF will no longer
be active during the next LIGO science run expected to begin in 2014,
developing and polishing the real-time capability will be highly
beneficial for synergy between LIGO and future successors of PTF, as
well as between PTF and current neutrino experiments such as Ice Cube.

To summarize, we have described the PTF operational setup for
real-time object detection and rapid multiwavelength follow-up
observations.  As an example, we have presented the real-time
discovery of the optical transient PTF10vdl by the PTF survey, and
results from a rapid-response multiwavelength follow-up campaign. Our
observations cover the very early phase (first days) following the
explosion of an underluminous Type II core-collapse SN, probably
discovered within 24\,hr of explosion, including very early
spectra. Our results demonstrate that the PTF wide-field optical
survey, in synergy with ToO programs using space- and ground-based
facilities, now enables us to conduct rapid multiwavelength
investigations of optical transients in a manner previously carried
out only for high-energy transients (GRBs). Our observations indicate
that we should expect observational breakthroughs in the study of SNe
and other optical transients similar to those recently achieved for
GRBs.

\section*{Acknowledgments}

This paper is dedicated to the memory of our dear colleague,
J. Jacobsen, who contributed significantly to the success of this
project.

The Palomar Transient Factory project is a scientific collaboration
between the California Institute of Technology, Columbia University,
Las Cumbres Observatory, the
Lawrence Berkeley National Laboratory, the National Energy Research
Scientific Computing Center, the University of Oxford, and the
Weizmann Institute of Science. Weizmann Institute participation in PTF
is supported in part by grants from the Israeli Science Foundation
(ISF) to A.G. Joint Weizmann-Caltech activity is supported by a grant
from the Binational Science Foundation to A.G. and S.R.K. Support for
Weizmann-UK collaborative work is provided by a Weizmann-UK ``making
connections'' grant to A.G. and M.S. Joint activity by A.G. and P.A.M.
is supported by a Weizmann-MINERVA grant. A.G. further acknowledges
support from an EU/FP7 Marie Curie IRG Fellowship. E.O.O. and D.P.
are grateful to NASA for Einstein Fellowships. The work of A.V.F.'s
group at UC Berkeley is funded by US National Science Foundation (NSF)
grant AST--0908886, the TABASGO Foundation, Gary and Cynthia Bengier,
and the Richard and Rhoda Goldman Fund.

LAIWO, a wide-angle camera operating on the 1\,m telescope at the Wise
Observatory, Israel, was built at the Max Planck Institute for
Astronomy (MPIA) in Heidelberg, Germany, with financial support from
the MPIA, grants from the German Israeli Science Foundation for
Research and Development, and the ISF. KAIT and its ongoing operation
were made possible by donations from Sun Microsystems, Inc., the
Hewlett-Packard Company, AutoScope Corporation, Lick Observatory, the
NSF, the University of California, the Sylvia \& Jim Katzman
Foundation, and the TABASGO Foundation.  The National Energy Research
Scientific Computing Center, which is supported by the Office of
Science of the U.S. Department of Energy under Contract No.
DE-AC02-05CH11231, provided staff, computational resources, and data
storage for this project. P.E.N. acknowledges support from the US
Department of Energy Scientific Discovery through Advanced Computing
program under contract DE-FG02-06ER06-04.

Some of the data presented herein were obtained at the W.~M. Keck
Observatory, which is operated as a scientific partnership among the
California Institute of Technology, the University of California, and
NASA; the observatory was made possible by the generous financial
support of the W. M. Keck Foundation. We thank the staffs of the many
observatories at which data were obtained for their excellent
assistance.  This research has made use of the NASA/IPAC Extragalactic
Database (NED), which is operated by the Jet Propulsion Laboratory,
California Institute of Technology, under contract with NASA.

\begin{deluxetable}{lllcc}
\tablecolumns{5}
\tablecaption{PTF10vdl: multicolor Photometry}
\tablehead{\colhead{Julian Date}  &  \colhead{Magnitude}  &  \colhead{Error}
 &  \colhead{Band}  &  \colhead{Telescope}\\
\colhead{[day]}  &  \colhead{[mag]}  &  \colhead{[mag]}
 &  & }
\startdata
2455450.77    &    21.34    &    99.99    &    r    &    P48 \\ 
2455450.82    &    21.47    &    99.99    &    r    &    P48 \\ 
2455459.13    &    18.86    &    0.10    &    r    &    P60 \\ 
2455460.16    &    18.94    &    0.05    &    r    &    P60 \\ 
2455463.12    &    18.80    &    0.07    &    r    &    P60 \\ 
2455468.40    &    18.65    &    0.05    &    r    &    P60 \\ 
2455468.41    &    18.73    &    0.07    &    r    &    P60 \\ 
2455469.19    &    18.71    &    0.04    &    r    &    P60 \\ 
2455470.38    &    18.85    &    0.03    &    r    &    P60 \\ 
2455473.33    &    18.81    &    0.03    &    r    &    P60 \\ 
2455477.35    &    18.88    &    0.02    &    r    &    P60 \\ 
2455478.15    &    18.83    &    0.03    &    r    &    P60 \\ 
2455479.12    &    18.85    &    0.02    &    r    &    P60 \\ 
2455481.15    &    18.84    &    0.01    &    r    &    P60 \\ 
2455483.15    &    18.87    &    0.03    &    r    &    P60 \\ 
2455484.20    &    18.86    &    0.02    &    r    &    P60 \\ 
2455486.14    &    18.77    &    0.04    &    r    &    P60 \\ 
2455496.30    &    18.85    &    0.08    &    r    &    P60 \\ 
2455497.15    &    18.82    &    0.02    &    r    &    P60 \\ 
2455499.14    &    18.79    &    0.03    &    r    &    P60 \\ 
2455501.10    &    18.82    &    0.03    &    r    &    P60 \\ 
2455503.15    &    18.81    &    0.03    &    r    &    P60 \\ 
2455507.25    &    18.73    &    0.06    &    r    &    P60 \\ 
2455459.13    &    18.55    &    0.10    &    i    &    P60 \\ 
2455460.16    &    18.87    &    0.05    &    i    &    P60 \\ 
2455463.12    &    18.72    &    0.09    &    i    &    P60 \\ 
2455468.40    &    18.58    &    0.06    &    i    &    P60 \\ 
2455468.41    &    18.81    &    0.08    &    i    &    P60 \\ 
2455469.19    &    18.66    &    0.05    &    i    &    P60 \\ 
2455470.38    &    18.69    &    0.03    &    i    &    P60 \\ 
2455473.33    &    18.63    &    0.03    &    i    &    P60 \\ 
2455477.35    &    18.70    &    0.03    &    i    &    P60 \\ 
2455478.15    &    18.60    &    0.03    &    i    &    P60 \\ 
2455481.15    &    18.68    &    0.05    &    i    &    P60 \\ 
2455482.12    &    18.61    &    0.04    &    i    &    P60 \\ 
2455483.14    &    18.60    &    0.03    &    i    &    P60 \\ 
2455484.19    &    18.58    &    0.04    &    i    &    P60 \\ 
2455485.12    &    18.57    &    0.04    &    i    &    P60 \\ 
2455486.12    &    18.59    &    0.06    &    i    &    P60 \\ 
2455496.30    &    18.58    &    0.06    &    i    &    P60 \\ 
2455497.15    &    18.49    &    0.03    &    i    &    P60 \\ 
2455499.14    &    18.45    &    0.05    &    i    &    P60 \\ 
2455501.10    &    18.52    &    0.03    &    i    &    P60 \\ 
2455503.14    &    18.44    &    0.04    &    i    &    P60 \\ 
2455507.25    &    18.41    &    0.05    &    i    &    P60 \\ 
2455459.13    &    18.96    &    0.19    &    g    &    P60 \\ 
2455460.17    &    18.93    &    0.05    &    g    &    P60 \\ 
2455463.13    &    18.97    &    0.11    &    g    &    P60 \\ 
2455468.41    &    18.96    &    0.07    &    g    &    P60 \\ 
2455468.41    &    18.91    &    0.08    &    g    &    P60 \\ 
2455469.20    &    18.95    &    0.04    &    g    &    P60 \\ 
2455470.39    &    19.04    &    0.04    &    g    &    P60 \\ 
2455473.35    &    19.10    &    0.04    &    g    &    P60 \\ 
2455477.35    &    19.28    &    0.02    &    g    &    P60 \\ 
2455478.15    &    19.28    &    0.02    &    g    &    P60 \\ 
2455479.13    &    19.27    &    0.02    &    g    &    P60 \\ 
2455480.18    &    19.32    &    0.03    &    g    &    P60 \\ 
2455481.15    &    19.29    &    0.02    &    g    &    P60 \\ 
2455482.13    &    19.31    &    0.03    &    g    &    P60 \\ 
2455483.15    &    19.33    &    0.03    &    g    &    P60 \\ 
2455484.20    &    19.31    &    0.03    &    g    &    P60 \\ 
2455485.12    &    19.33    &    0.03    &    g    &    P60 \\ 
2455486.14    &    19.28    &    0.07    &    g    &    P60 \\ 
2455496.31    &    19.49    &    0.09    &    g    &    P60 \\ 
2455497.15    &    19.43    &    0.03    &    g    &    P60 \\ 
2455499.14    &    19.43    &    0.04    &    g    &    P60 \\ 
2455503.15    &    19.51    &    0.04    &    g    &    P60 \\ 
2455507.25    &    19.42    &    0.07    &    g    &    P60 \\ 
2455456.93    &    19.00    &    0.12    &    r    &    LT \\ 
2455458.94    &    18.86    &    0.04    &    r    &    LT \\ 
2455467.97    &    18.81    &    0.06    &    r    &    LT \\ 
2455470.94    &    18.84    &    0.12    &    r    &    LT \\ 
2455472.97    &    18.79    &    0.11    &    r    &    LT \\ 
2455473.90    &    18.80    &    0.03    &    r    &    LT \\ 
2455474.89    &    18.85    &    0.11    &    r    &    LT \\ 
2455475.88    &    18.82    &    0.03    &    r    &    LT \\ 
2455478.91    &    18.81    &    0.05    &    r    &    LT \\ 
2455481.94    &    18.83    &    0.05    &    r    &    LT \\ 
2455483.96    &    18.78    &    0.03    &    r    &    LT \\ 
2455485.85    &    18.80    &    0.03    &    r    &    LT \\ 
2455491.03    &    18.79    &    0.05    &    r    &    LT \\ 
2455491.93    &    18.83    &    0.09    &    r    &    LT \\ 
2455496.89    &    18.75    &    0.02    &    r    &    LT \\ 
2455497.82    &    18.84    &    0.23    &    r    &    LT \\ 
2455498.84    &    18.85    &    0.14    &    r    &    LT \\ 
2455500.85    &    18.83    &    0.11    &    r    &    LT \\ 
2455505.88    &    18.85    &    0.11    &    r    &    LT \\ 
2455506.91    &    18.81    &    0.04    &    r    &    LT \\ 
2455512.88    &    18.83    &    0.05    &    r    &    LT \\ 
2455518.84    &    18.85    &    0.04    &    r    &    LT \\ 
2455456.93    &    18.98    &    0.18    &    g    &    LT \\ 
2455458.94    &    18.90    &    0.04    &    g    &    LT \\ 
2455467.97    &    19.00    &    0.15    &    g    &    LT \\ 
2455470.94    &    19.12    &    0.25    &    g    &    LT \\ 
2455472.97    &    19.11    &    0.08    &    g    &    LT \\ 
2455473.90    &    19.18    &    0.04    &    g    &    LT \\ 
2455474.89    &    19.20    &    0.08    &    g    &    LT \\ 
2455475.88    &    19.20    &    0.07    &    g    &    LT \\ 
2455478.91    &    19.28    &    0.05    &    g    &    LT \\ 
2455481.93    &    19.27    &    0.03    &    g    &    LT \\ 
2455483.95    &    19.32    &    0.08    &    g    &    LT \\ 
2455485.85    &    19.34    &    0.04    &    g    &    LT \\ 
2455491.03    &    19.30    &    0.17    &    g    &    LT \\ 
2455491.93    &    19.41    &    0.13    &    g    &    LT \\ 
2455496.89    &    19.42    &    0.02    &    g    &    LT \\ 
2455497.82    &    19.47    &    0.18    &    g    &    LT \\ 
2455500.85    &    19.48    &    0.04    &    g    &    LT \\ 
2455501.85    &    19.49    &    0.18    &    g    &    LT \\ 
2455505.88    &    19.51    &    0.10    &    g    &    LT \\ 
2455506.91    &    19.49    &    0.03    &    g    &    LT \\ 
2455512.88    &    19.61    &    0.04    &    g    &    LT \\ 
2455518.84    &    19.72    &    0.19    &    g    &    LT \\ 
2455456.93    &    18.75    &    0.23    &    i    &    LT \\ 
2455458.94    &    18.71    &    0.03    &    i    &    LT \\ 
2455467.97    &    18.67    &    0.12    &    i    &    LT \\ 
2455470.94    &    18.66    &    0.26    &    i    &    LT \\ 
2455472.97    &    18.71    &    0.10    &    i    &    LT \\ 
2455473.90    &    18.57    &    0.03    &    i    &    LT \\ 
2455474.89    &    18.55    &    0.05    &    i    &    LT \\ 
2455475.88    &    18.64    &    0.08    &    i    &    LT \\ 
2455478.91    &    18.60    &    0.03    &    i    &    LT \\ 
2455479.86    &    18.66    &    0.11    &    i    &    LT \\ 
2455481.94    &    18.58    &    0.08    &    i    &    LT \\ 
2455483.96    &    18.64    &    0.13    &    i    &    LT \\ 
2455485.85    &    18.54    &    0.03    &    i    &    LT \\ 
2455491.04    &    18.51    &    0.04    &    i    &    LT \\ 
2455491.94    &    18.51    &    0.04    &    i    &    LT \\ 
2455496.89    &    18.45    &    0.02    &    i    &    LT \\ 
2455500.85    &    18.56    &    0.25    &    i    &    LT \\ 
2455501.86    &    18.47    &    0.24    &    i    &    LT \\ 
2455505.88    &    18.64    &    0.18    &    i    &    LT \\ 
2455506.91    &    18.48    &    0.05    &    i    &    LT \\ 
2455512.88    &    18.50    &    0.06    &    i    &    LT \\ 
2455518.84    &    18.44    &    0.04    &    i    &    LT \\ 
2455457.34    &    18.99    &    0.10    &    r    &    LCOGT \\ 
2455467.58    &    18.78    &    0.05    &    r    &    LCOGT \\ 
2455468.60    &    18.89    &    0.25    &    r    &    LCOGT \\ 
2455475.60    &    18.93    &    0.20    &    r    &    LCOGT \\ 
2455484.34    &    18.83    &    0.01    &    r    &    LCOGT \\ 
2455496.34    &    18.87    &    0.24    &    r    &    LCOGT \\ 
2455468.60    &    17.57    &    0.33    &    g    &    LCOGT \\ 
2455480.47    &    17.83    &    0.26    &    g    &    LCOGT \\ 
2455493.42    &    18.07    &    0.24    &    g    &    LCOGT \\ 
2455496.34    &    17.97    &    0.34    &    g    &    LCOGT \\ 
2455454.81    &    19.36    &    0.23    &    R    &    Wise  \\ 
2455451.25    &    19.61    &    99.99   &    Clear    &    KAIT  \\ 
2455455.33    &    18.98    &    0.09    &    Clear    &    KAIT  \\ 
2455456.40    &    19.01    &    0.06    &    Clear    &    KAIT  \\ 
2455458.33    &    18.90    &    0.09    &    Clear    &    KAIT  \\ 
2455464.35    &    18.83    &    0.14    &    Clear    &    KAIT  \\ 
2455466.34    &    18.79    &    0.11    &    Clear    &    KAIT  \\ 
2455470.30    &    18.75    &    0.12    &    Clear    &    KAIT  \\ 
2455479.23    &    18.93    &    0.08    &    Clear    &    KAIT  \\ 
2455482.27    &    18.86    &    0.11    &    Clear    &    KAIT  \\ 
2455519.12    &    19.10    &    0.14    &    Clear    &    KAIT  \\ 
2455529.17    &    18.84    &    0.13    &    Clear    &    KAIT  \\ 
2455456.31    &    19.00    &    0.06    &    R    &    KAIT  \\ 
2455458.28    &    19.07    &    0.08    &    R    &    KAIT  \\ 
2455464.33    &    18.99    &    0.07    &    R    &    KAIT  \\ 
2455466.31    &    18.78    &    0.06    &    R    &    KAIT  \\ 
2455468.30    &    18.82    &    0.06    &    R    &    KAIT  \\ 
2455470.29    &    18.86    &    0.06    &    R    &    KAIT  \\ 
2455476.25    &    18.83    &    0.07    &    R    &    KAIT  \\ 
2455479.21    &    18.86    &    0.04    &    R    &    KAIT  \\ 
2455482.23    &    18.81    &    0.05    &    R    &    KAIT  \\ 
2455485.20    &    18.70    &    0.07    &    R    &    KAIT  \\ 
2455519.16    &    18.66    &    0.21    &    R    &    KAIT  \\ 
2455529.13    &    18.85    &    0.06    &    R    &    KAIT  \\ 
2455532.12    &    18.92    &    0.05    &    R    &    KAIT  \\ 
2455537.12    &    18.89    &    0.07    &    R    &    KAIT  \\ 
2455540.14    &    18.90    &    0.09    &    R    &    KAIT  \\ 
2455546.12    &    19.05    &    0.14    &    R    &    KAIT  \\ 
2455456.30    &    19.21    &    0.07    &    B    &    KAIT  \\ 
2455458.27    &    19.18    &    0.15    &    B    &    KAIT  \\ 
2455464.32    &    19.28    &    0.16    &    B    &    KAIT  \\ 
2455466.30    &    19.55    &    0.28    &    B    &    KAIT  \\ 
2455468.29    &    19.54    &    0.08    &    B    &    KAIT  \\ 
2455470.28    &    19.57    &    0.11    &    B    &    KAIT  \\ 
2455473.25    &    19.72    &    0.11    &    B    &    KAIT  \\ 
2455476.24    &    19.67    &    0.08    &    B    &    KAIT  \\ 
2455479.21    &    19.89    &    0.09    &    B    &    KAIT  \\ 
2455482.22    &    19.96    &    0.13    &    B    &    KAIT  \\ 
2455485.19    &    20.03    &    0.22    &    B    &    KAIT  \\ 
2455519.15    &    20.30    &    0.24    &    B    &    KAIT  \\ 
2455529.12    &    20.37    &    0.15    &    B    &    KAIT  \\ 
2455532.11    &    20.44    &    0.14    &    B    &    KAIT  \\ 
2455537.11    &    20.48    &    0.16    &    B    &    KAIT  \\ 
2455456.31    &    18.98    &    0.06    &    V    &    KAIT  \\ 
2455458.28    &    18.84    &    0.11    &    V    &    KAIT  \\ 
2455464.33    &    18.93    &    0.09    &    V    &    KAIT  \\ 
2455466.30    &    18.88    &    0.07    &    V    &    KAIT  \\ 
2455468.29    &    18.81    &    0.07    &    V    &    KAIT  \\ 
2455470.29    &    18.94    &    0.08    &    V    &    KAIT  \\ 
2455473.25    &    18.94    &    0.07    &    V    &    KAIT  \\ 
2455476.25    &    19.01    &    0.06    &    V    &    KAIT  \\ 
2455479.21    &    19.09    &    0.05    &    V    &    KAIT  \\ 
2455482.22    &    19.00    &    0.06    &    V    &    KAIT  \\ 
2455485.19    &    18.77    &    0.15    &    V    &    KAIT  \\ 
2455529.13    &    19.37    &    0.08    &    V    &    KAIT  \\ 
2455532.12    &    19.23    &    0.06    &    V    &    KAIT  \\ 
2455537.12    &    19.35    &    0.11    &    V    &    KAIT  \\ 
2455540.13    &    19.12    &    0.10    &    V    &    KAIT  \\ 
2455546.12    &    19.17    &    0.15    &    V    &    KAIT  \\ 
2455456.32    &    19.22    &    0.22    &    I    &    KAIT  \\ 
2455458.29    &    18.62    &    0.13    &    I    &    KAIT  \\ 
2455464.34    &    18.37    &    0.09    &    I    &    KAIT  \\ 
2455466.31    &    18.42    &    0.10    &    I    &    KAIT  \\ 
2455468.30    &    18.46    &    0.09    &    I    &    KAIT  \\ 
2455470.29    &    18.51    &    0.10    &    I    &    KAIT  \\ 
2455476.25    &    18.49    &    0.12    &    I    &    KAIT  \\ 
2455479.22    &    18.25    &    0.07    &    I    &    KAIT  \\ 
2455482.23    &    18.32    &    0.08    &    I    &    KAIT  \\ 
2455485.20    &    18.26    &    0.14    &    I    &    KAIT  \\ 
2455529.13    &    17.97    &    0.08    &    I    &    KAIT  \\ 
2455532.12    &    17.94    &    0.08    &    I    &    KAIT  \\ 
2455537.13    &    18.06    &    0.07    &    I    &    KAIT  \\ 
2455540.14    &    18.13    &    0.10    &    I    &    KAIT  \\ 
2455454.77    &    19.70    &    0.08    &    r    &    P48 \\ 
2455458.80    &    19.04    &    0.22    &    r    &    P48 \\ 
2455464.76    &    18.99    &    0.04    &    r    &    P48 \\ 
2455465.82    &    18.83    &    0.17    &    r    &    P48 \\ 
2455466.71    &    18.85    &    0.01    &    r    &    P48 \\ 
2455472.77    &    18.83    &    0.08    &    r    &    P48 \\ 
2455473.76    &    18.83    &    0.10    &    r    &    P48 \\ 
2455481.66    &    18.85    &    0.02    &    r    &    P48 \\ 
2455482.65    &    18.84    &    0.09    &    r    &    P48 \\ 
2455457.29    &    17.46    &    0.07    &     U     &     Swift   \\ 
2455457.87    &    17.63    &    0.08    &     U     &     Swift   \\ 
2455458.71    &    17.73    &    0.07    &     U     &     Swift   \\ 
2455459.34    &    17.78    &    0.17    &     U     &     Swift   \\ 
2455459.41    &    17.66    &    0.13    &     U     &     Swift   \\ 
2455459.48    &    17.84    &    0.15    &     U     &     Swift   \\ 
2455463.73    &    18.06    &    0.09    &     U     &     Swift   \\ 
2455466.27    &    18.21    &    0.12    &     U     &     Swift   \\ 
2455457.29    &    17.47    &    0.07    &     UVW1     &     Swift   \\ 
2455457.88    &    17.49    &    0.07    &     UVW1     &     Swift   \\ 
2455458.72    &    17.73    &    0.07    &     UVW1     &     Swift   \\ 
2455459.41    &    17.72    &    0.12    &     UVW1     &     Swift   \\ 
2455459.48    &    17.84    &    0.13    &     UVW1     &     Swift   \\ 
2455463.72    &    18.70    &    0.12    &     UVW1     &     Swift   \\ 
2455466.27    &    19.63    &    0.27    &     UVW1     &     Swift   \\ 
2455457.29    &    17.40    &    0.07    &     UVM2     &     Swift   \\ 
2455457.87    &    17.45    &    0.07    &     UVM2     &     Swift   \\ 
2455458.70    &    17.88    &    0.08    &     UVM2     &     Swift   \\ 
2455459.34    &    17.80    &    0.16    &     UVM2     &     Swift   \\ 
2455459.41    &    17.76    &    0.12    &     UVM2     &     Swift   \\ 
2455459.48    &    18.07    &    0.15    &     UVM2     &     Swift   \\ 
2455463.72    &    18.99    &    0.15    &     UVM2     &     Swift   \\ 
2455466.27    &    20.18    &    0.38    &     UVM2     &     Swift   \\ 
2455457.39    &    17.64    &    0.07    &     UVW2     &     Swift   \\ 
2455457.87    &    17.80    &    0.07    &     UVW2     &     Swift   \\ 
2455458.72    &    17.98    &    0.05    &     UVW2     &     Swift   \\ 
2455459.34    &    18.11    &    0.13    &     UVW2     &     Swift   \\ 
2455459.41    &    18.03    &    0.10    &     UVW2     &     Swift   \\ 
2455459.48    &    18.08    &    0.11    &     UVW2     &     Swift   \\ 
2455463.72    &    18.99    &    0.11    &     UVW2     &     Swift   \\ 
2455466.26    &    19.99    &    0.25    &     UVW2     &     Swift   \\ 
\enddata
\tablecomments{Multiband photometry of PTF10vdl. All optical ground-based
data are calibrated onto the SDSS grid. {\it Swift} data are calibrated 
following Poole et al. (2008). The P60, LT and Wise $gri$ data calibrate
consistently onto the SDSS grid, while small constant offsets had
to be applied to the LCOGT, P48 and KAIT data; values reported here
include these offsets. Values of 99.99 in the error column indicate that
the values in the magnitude column are upper limits.} 
\end{deluxetable}

\end{document}